\documentclass[preprint,12pt]{elsarticle}

\usepackage{amssymb}
\usepackage{graphicx}
\usepackage{dcolumn}
\usepackage{bm}
\usepackage{color}

\journal{Current Applied Physics}

\begin{document}

\begin{frontmatter}

\title{Evidence of magnetic field quenching of phosphorous-doped silicon quantum dots}

\author[1]{M. F. Gonzalez-Zalba}
 \ead{mg507@cam.ac.uk}
\author[2]{J. Galibert}
\author[2]{F. Iacovella}
\author[1]{D. Williams}
\author[1]{T. Ferrus}

\address[1]{Hitachi Cambridge Laboratory, J. J. Thomson Avenue, CB3 0HE, Cambridge, \\ United Kingdom}
\address[2]{Laboratoire National des Champs Magn\'etiques Intenses, 143 Avenue de Rangueil, 31500 Toulouse, France}

\begin{abstract}

We present data on the electrical transport properties of highly-doped silicon-on-insulator quantum dots under the effect of pulsed magnetic fields up to 48 T. At low field intensities, $B<7$ T, we observe a strong modification of the conductance due to the destruction of weak localization whereas at higher fields, where the magnetic field length becomes comparable to the effective Bohr radius of phosphorous in silicon, a strong decrease in conductance is demonstrated. Data in the high and low electric field bias regimes are then compared to show that close to the Coulomb blockade edge magnetically-induced quenching to single donors in the quantum dot is achieved at about 40\,T.

\end{abstract}


\end{frontmatter}

\section{Introduction.}

Confined electron spins in semiconductor materials provide one of the most promising candidates for a physical implementation of quantum information processing.

One of the most successful approach to confine an electron into a single, double or triple quantum dots \cite{Hanson, VanderWiel, Vandersypen} remains the use of the well-known split-gate technique \cite{Split gate} in GaAs-based heterostructures. Achieving the single electron regime in silicon is technologically far more challenging due to the larger effective mass, interface disorder and electron localization effects, but the promise of long lifetimes and coherence times due to almost nuclear-free and low spin-orbit coupling environment makes that endeavor worthwhile. Despite the increased fabrication complexity, single electron spin have been electrostatically isolated in single and double quantum dots in SiGe heterostructures \cite{SimmonsSiGe, Maune} as well as in metal-oxide-semiconductor silicon quantum dots \cite{Yang, Horibe}.

With the recent progress in single ion implantation \cite{Jamieson} and atomically-precise hydrogen-resist lithography \cite{SimmonsSTM}, the development and study of single atom devices \cite{Dupont, Gonzalez-Zalba} became possible which led to a new way of implementing a solid-state quantum computer \cite{Kane, Pla} or realizing high-frequency single electron pumps \cite{Roche}. Here, the electronic confinement is achieved by the sole Coulomb potential of the impurity and at sufficiently low temperatures an electron spin is localized to the positively-charged nucleus. However, in order to observe single dopant signatures in the electronic transport auxiliary electric field are still necessary.

Alternatively, high magnetic fields provide a way to modify the electronic properties of donor electrons. For example, the application of 8.5 T in bulk Si:P system was shown to improve the coherence of the readout mechanism \cite{Morley, Boehme} and at 33 T strong spatial modification of the atomic wavefunctions was achieved \cite{Murdin}. However, so far, no study has focused on magnetically-induced electron confinement to single donors.

Because electron localization can also be influenced by mean of a perpendicular magnetic field, magnetically compressing doped silicon quantum dots should be possible and, at sufficiently high magnetic field, single dopant transport could be observed. A change in transport behavior is thus expected when the magnetic field length becomes comparable to the effective Bohr radius of the impurity.

Here, we present a preliminary study on such an effect in a highly-doped silicon on insulator (SOI) quantum dot subjected to pulsed magnetic fields up to 48\,T. First, by observing a magnetic field induced metal-to-insulator transition, we demonstrate that magnetic field mostly affects a narrow region around the quantum dot and that large sections of the device, including leads are only weakly affected. Also we suggest that one of the effect of magnetic field is to reduce the effective dot diameter. We then describe the three regimes observed : the destruction of weak localization at low field ($B < 7$ T), an intermediate region where the magnetoconductivity becomes non-monotonous ($7< B <40$ T) and, at high fields ($B > 40$ T), a sharp decrease in conductivity with field that we associate with quantum dot quenching around single donors. In section 2, we describe the device as well as the experiment techniques of pulsed fields. We also discuss the general characteristics in gate and source-drain bias as well as the behavior of the conductivity in temperature in the absence of a field. Section 3 concerns magnetic field experiments, first at high bias (Sec. 3.1), then at low bias (Sec. 3.2). The section concludes with a study of the variation of conductivity with field at the lowest temperature for different biases. Discussion and conclusions are given in section 4.

\section{Device characteristics and experimental technique.}

\begin{figure}
\begin{center}
\includegraphics[width=85mm, bb=4 2 254 276]{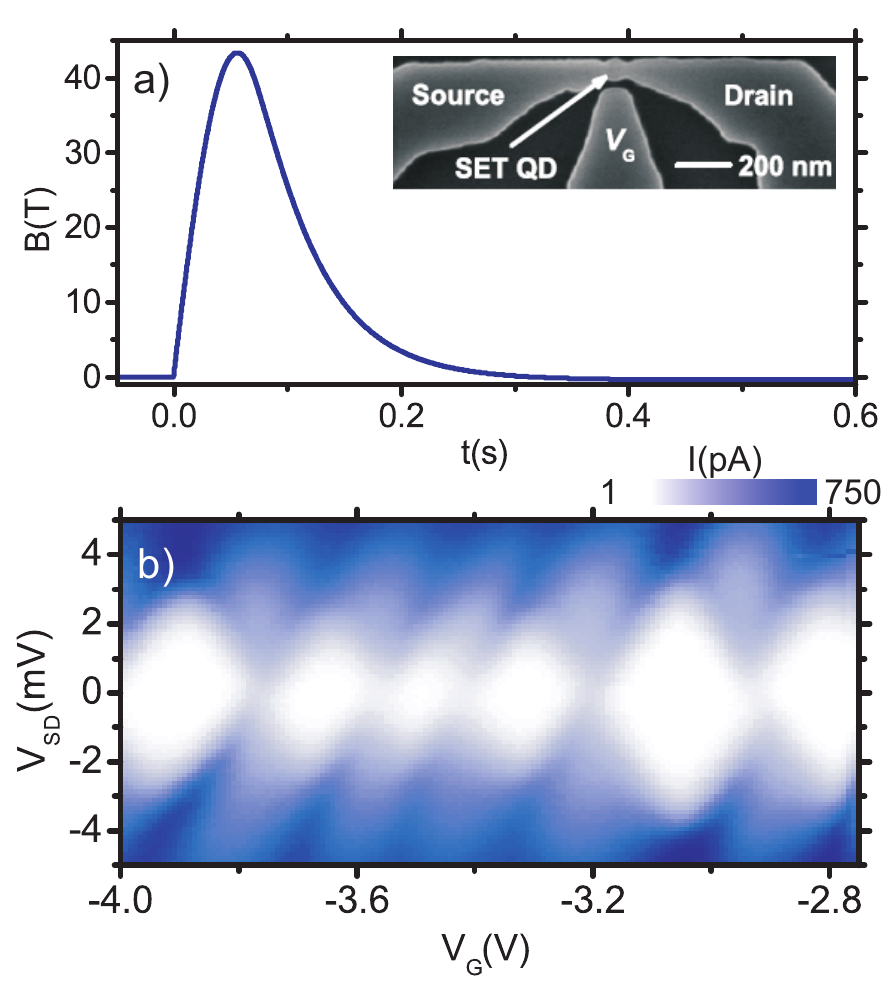}
\end{center}
\caption{\label{fig:figure0} \textbf{a}, Time dependence of the magnetic field during a pulse. Inset shows a SEM image of a device indicating the source, drain and gate contacts. \textbf{b} Coulomb diamonds obtained at 1.8\,K and in the absence of magnetic field.}
\end{figure}

Devices were fabricated from a 40 nm-thick SOI wafer by a single phosphorous implant through a 20\,nm protective silicon oxide. This provided a dopant density of $N_{\textup{\tiny{D}}} \sim 3\times 10^{19}$\,cm$^{-3}$ after thermal activation. A single dot of diameter $\sim$60\,nm diameter as well as a side gate, the source and the drain contacts were geometrically defined by electron beam lithography and reactive ion etching in a single process stage. After the implant, the sacrificial oxide was removed and defects were passivated by growing a 15\,nm thermal oxide.

The tunnel barriers are generated at the constrictions of the device due to the formation of a localizing region at the periphery of the quantum dot \cite{Ferrus}. This region is of variable size depending on the local electric field that is controlled by the side gate. The presence of an Si-SiO$_2$ interface where phosphorous atoms are diffusing towards during the oxidation process contributes largely to the localization process. Due to that phenomenon, devices show an insulating behavior at low temperature and at low source-drain bias $V_{\textup{\tiny{SD}}}$ despite the high doping concentration. However, for $eV_{\textup{\tiny{SD}}}\gg E_{\textup{\tiny{C}}}$ and in the absence of magnetic field, transport is dominated by conduction in the leads.

Measurements were performed in a variable temperature cryostat between 1.8 K and 200 K. The gate and source bias were provided by two Hewlett Packard 3245A whereas the current measurement was obtained by amplifying the current from the drain by the use of a Princeton Applied Research 181 current-to-voltage amplifier and then reading the output voltage with a Hewlett Packard 3458A. High pulsed magnetic fields were produced by first charging a 5 MJ capacitor bank with a 20 kV voltage and then discharging the energy into the magnet coil. This produces a magnetic field pulse of duration 400 ms and of maximum field of 48 T (Fig. 1a). The exact field strength was measured directly from the current intensity flowing through the magnet as well as from the a pick-up coil located in close proximity of the device. In all experiments the field orientation was perpendicular to the SOI plane.

Devices were first characterized at 1.8 K in the absence of magnetic field. They all show clear Coulomb diamonds but of variable size (Fig. 1b). The average charging energy $E_{\textup{\tiny{C}}}$ is estimated to be 2 meV, consistent with a quantum dot diameter of about 60 nm as measured from Scanning Electron Microscope image (SEM) (Inset Fig 1a). In all measurements the gate voltage was kept to $V_{\textup{\tiny{G}}}$=0\,V.

\section{Magnetic field}

\subsection {High bias regime : Metal to insulator transition and magnetoresistance.}

\begin{figure}
\begin{center}
\includegraphics[width=85mm, bb=0 11 255 327]{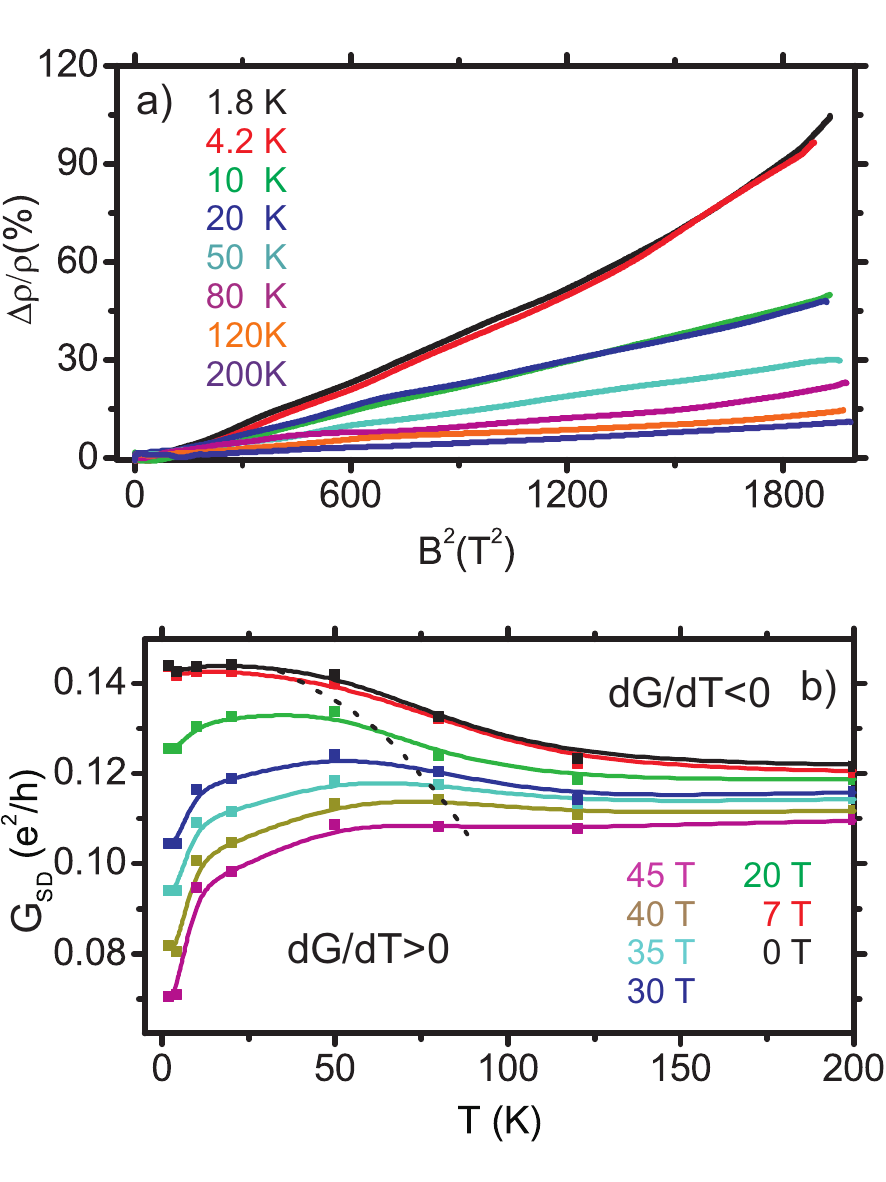}
\end{center}
\caption{\label{fig:figure1} \textbf{a}, Magnetoresistance at different temperatures measured at $V_{\textup{\tiny{SD}}}$ = 40 mV. \textbf{b}, Source-drain conductance as a function of temperature and field mapping the metal-to-insulator transition at $V_{\textup{\tiny{SD}}} = 40$ mV. The dotted line shows the expected positions in temperature and magnetic field of the conductivity maximum as given by Eq. 2.}
\end{figure}

For $eV_{\textup{\tiny{SD}}}\gg E_{\textup{\tiny{C}}}$  we observed a nearly quadratic increase in magnetoresistance, for a wide range of temperatures despite significant deviations below 4.2 K (Fig. 2a). This behavior is expected for both metal and semiconductors where the free electron Fermi surface is closed \cite{Porter}. However, we do not observe any saturation even at high field, both at low and high temperatures. Such a situation can occur if some of the orbits are open. In this case, the conductivity tensor includes an extra contribution from small angle scattering events that remove the saturation at high field \cite{Pippard}. Such open orbits are commonly found in disorder systems \cite{Littlewood}. In our device, randomness in dopant distribution, localized centers at the Si-SiO$_2$ interface may result in substantial scattering, so that the low field regime determined by $\omega_{\textup{\tiny{C}}} \tau \ll 1$, where $\omega_{\textup{\tiny{C}}}$ is the cyclotron frequency and $\tau$ the mean scattering time, is always observed in practice. 

Additional information on the influence of dopants in this structure is revealed by the temperature dependence of the conductivity $G_{\textup{\tiny{SD}}} \left(T \right)$ at different magnetic fields (Fig. 2b). At $B=0$ T, the conductivity $G_{\textup{\tiny{SD}}}$ is seen to increase when lowering the temperature until about 6\, K, below which it starts decreasing weakly. This general behavior is maintained while increasing the magnetic field. However, in the region where $\textup{d}G_{\textup{\tiny{SD}}}/\textup{d}T < 0$, the high temperature value of $G_{\textup{\tiny{SD}}}$ slightly decreases with $B$ and its variation in temperature becomes weaker, until $\textup{d}G_{\textup{\tiny{SD}}}/\textup{d}T \sim 0$ at about 40\,T. Such a behavior is similar to the one observed in metals or in semiconductors deep into the metallic regime. Because low-dimensional structures (below 2D) can only exist in the localized regime \cite{Abrahams} and, owing to both the large dimensions of the leads and their high doping concentration, the metallic behavior of the conductivity in that region can be ascribed to the one of the largest sections of the device, including source and drain leads.

Below a critical temperature $T_{\textup{\tiny{C}}}$ that increases with field, a regime where $\textup{d}G_{\textup{\tiny{SD}}}/\textup{d}T > 0$ is observed. Such a change in the sign of $\textup{d}G_{\textup{\tiny{SD}}}/\textup{d}T$ can be accounted for a competition between the metallicity of the lead regions and the insulating characteristic of the section of the device around the quantum dot. This includes the constriction regions as well as the dot itself. Indeed charging effects are negligible as long as $e V_{\textup{\tiny{SD}}} , k_{\textup{\tiny{B}}}T \gg E_{\textup{\tiny{C}}}$ which is the case at $V_{\textup{\tiny{SD}}} = 40$ mV. The magnetoresistance being positive at all temperatures, we can assume that most of the magnetic field effect is indeed coming from an increase in electron localization in the quantum dot. It is interesting to notice that the extent of the depletion layer in these quantum dot do not extend further than 20 nm from the Si-SiO$_2$ interface, as estimated from DC measurements at zero field \cite{Ferrus}. In order to decrease significantly the conductivity as observed at 45 T, it is thus necessary that the magnetic field enhances the depletion so that the effective dot diameter gets controlled by the magnetic field length $l_{\textup{\tiny{B}}} = \left( \hbar /eB\right)^{1/2}$. If $T_{\textup{\tiny{C}}}$ is defined by the temperature below which Coulomb blockade is observable, its variation with field can then easily be approximate by the following equation :

\begin{eqnarray}\label{eqn:equation1}
4.35 k_{\textup{\tiny{B}}} T = \frac{e^2}{4\pi \epsilon_{\textup{\tiny{Si}}} l_{\textup{\tiny{B}}}} 
\end{eqnarray}

\begin{eqnarray}\label{eqn:equation2}
T_{\textup{\tiny{C}}} = \left( \frac{B}{B_0} \right)^{1/2} \textup{with}\, B_0 \approx  12.77 \ \textup{TK}^{-2}
\end{eqnarray}

where $\epsilon_{\textup{\tiny{Si}}} = 11.7$ is the silicon dielectric constant and the factor 4.35 results from multi-level tunneling in the dot.

This is in reasonably good agreement with observations (Fig. 2b). Finally, in order to obtain further confirmation of our explanation we study the low bias limit at 4.2\,K where the transport is expected to be dominated by electron tunneling through the quantum dot over the source and drain leads.

\subsection {Intermediate and low bias regime : Weak localization and donor quenching.}

\begin{figure}
\begin{center}
\includegraphics[width=85mm, bb=1 1 256 310]{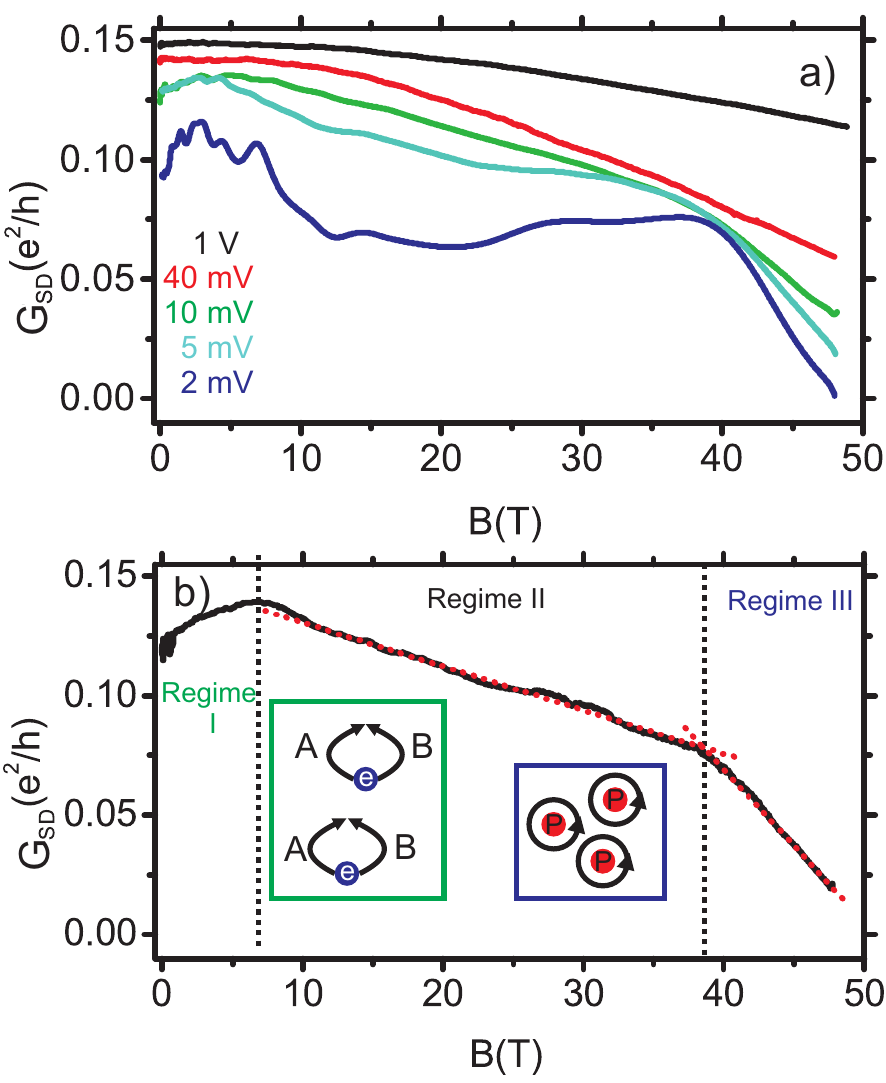}
\end{center}
\caption{\label{fig:figure3} \textbf{a} Source-drain conductance with magnetic field in the high and low bias limit at 4~K. \textbf{b} Conductance as a function of field for $V_{\textup{\tiny{SD}}}$=7.5~mV. The three observed regimes : (I) weak localization at low field, (II) transport in non-homogeneous confined material and (III) field quenching of the quantum dot at higher field. Green inset represent the characteristic electron self-interference mechanism in weak localization. Navy insert represent electron localization around single donors.}
\end{figure}

When the source-drain bias is set in proximity of the edge of the blockade region,$V_{\textup{\tiny{SD}}}$ $\geq$ $E_{\textup{\tiny{C}}}$, the magnetoconductance looses its quadratic behavior. This is demonstrated in Fig. 3a where a comparison between the low ($V_{\textup{\tiny{SD}}}$ $\leq$10~mV) and high ($V_{\textup{\tiny{SD}}}$ $>$10~mV) bias conductance as a function of magnetic field is made. Moreover, we identify three new regimes appearing at low (I), intermediate (II) and high fields (III) that are exemplified in Fig. 3b where $V_{\textup{\tiny{SD}}}$=7.5~mV:

\begin{itemize}

  \item[(I)\,\,\,\,] Below 6 T, weak localization \cite{weak localization} is efficiently suppressed by magnetic field. This reduces backscattering events and consequently increases the conductivity until strong localization is attained which leads to a maximum in the conductivity. Weak localization itself may result from conduction via edge states in the quantum dot \cite{edge state}.

  \item[(II)\,\,] The second regime shows a linear variation with field up to 39 T. Such a variation has already been observed in inhomogeneous material. In our devices, this can be explained by the presence of a non-uniform doping density and the depletion layer near the Si-SiO$_2$ interface \cite{Porter, linear magneto}.

  \item[(III)] Above about 39 T, $G_{\textup{\tiny{SD}}} \left( B \right)$ remains roughly linear but with a slope approximately 3.5 times steeper than in the region II. In this range, $l_{\textup{\tiny{B}}}$ is comparable to $a_{\textup{\tiny{B}}}$ the effective Bohr radius in silicon. Because at high field $a_{\textup{\tiny{B}}}$ increases with $B$ \cite{Bohr}, the strong magnetic field regime is attained when $l_{\textup{\tiny{B}}} = a_{\textup{\tiny{B}}} \left( B \right) = 4.1$ nm $\sim 2 a_{\textup{\tiny{B}}} \left( 0 \right)$, if defining the effective Bohr radius in silicon from the Mott criterion $N_{\textup{\tiny{D}}}^{1/3} a_{\textup{\tiny{B}}} \sim 0.25$. At such fields, the magnetic field localizes electrons around single Phosphorous atoms, effectively reducing the tunneling rates in and out the dot. Quantum corrections to conductivity are expected to be important and there is a substantial shrinkage of the donor wavefunction that becomes entirely determined by the magnetic field rather than the localization length \cite{Wavefunction}.

\end{itemize}

By decreasing the bias so that $eV_{\textup{\tiny{SD}}} \approx E_{\textup{\tiny{C}}}$, the behaviors of $G_{\textup{\tiny{SD}}} \left( B \right)$ in the regions I and III remain unchanged despite the destruction of the weak localization being more efficient in region I and localization being enhanced in region III (Fig. 3a). This may arise from a reduction of inelastic tunneling at low bias. However, most the changes appear in the regime II where $G_{\textup{\tiny{SD}}} \left( B \right)$ becomes non monotonous with a clear minimum conductivity at about 20 T. This suggests an effect from Coulomb blockade and so, from confinement. It is also remarkable that the conductivity at the critical field of 39 T has little variation with $V_{\textup{\tiny{SD}}}$. 

This may be explained by considering that the dot confinement is driven by the magnetic field and that the charging energy increases with the field strength. Experiments being performed at constant source-drain bias, the conductivity is expected to fluctuate until $E_{\textup{\tiny{C}}} > eV_{\textup{\tiny{SD}}}$ and blockade is reached, which leads to an abrupt decrease in conductivity. This effect may be enhanced by an enlargement of the tunnel barriers with field and so, by a reduction of the tunneling rates.

\section{Conclusions.}

The application of a perpendicular and high magnetic field to highly doped silicon quantum dots has revealed the strong influence of localizing centers and scattering events on magneto-transport properties. In particular, a comparison with the low bias data reveal a sharp reduction in the conductance at 39~T showing the first evidence of magnetically-induced quenching to single donors in silicon quantum dots. More generally, these studies suggest that the depletion layer at the vicinity of the Si-SiO$_2$ interface that can be controlled by means of an electric field can also be significantly affected by a magnetic field, providing an alternative way to manipulate in-situ the quantum dot confinement. Although high magnetic fields introduces an extra level of complexity to already complex quantum information processing devices, this preliminary experiment proposes a route to probe electrically the properties of single impurity atoms under the additional confinement induced by magnetic fields without the need of devices that present single dopant signatures in their electrical transport properties.

\section{Acknowledgement}

This work was supported by Project for Developing Innovation Systems of the Ministry of Education, Culture, Sports, Science and Technology (MEXT). We also would like to thank Prof. Sir M. Pepper for discussions and feedback on the manuscript.

\end{document}